\documentclass[conference]{IEEEtran}
\IEEEoverridecommandlockouts

\usepackage{graphicx}
\usepackage{cite}  
\usepackage{amsmath,amssymb,amsfonts}
\usepackage{xcolor}
\usepackage{tabularx}
\usepackage{array}
\usepackage{url}
\usepackage{float}
\usepackage{stfloats} 
\usepackage{hyperref}  

\def\BibTeX{{\rm B\kern-.05em{\sc i\kern-.025em b}\kern-.08em
    T\kern-.1667em\lower.7ex\hbox{E}\kern-.125emX}}

\begin{document}
\renewcommand\thesection{\arabic{section}}
\renewcommand\thesubsection{\thesection.\arabic{subsection}}

\makeatletter
\renewcommand{\@seccntformat}[1]{\csname the#1\endcsname\quad}
\makeatother

\setcounter{secnumdepth}{2}

\title{System Security Framework for 5G Advanced /6G IoT Integrated Terrestrial Network-Non-Terrestrial Network (TN-NTN) with AI-Enabled Cloud Security }

\author{
\centering
\begin{tabular}{c}
Sasa Maric\textsuperscript{1}, Rasil Baidar\textsuperscript{2}, Robert Abbas\textsuperscript{2}, Sam Reisenfeld\textsuperscript{3} \\
\text{s.maric@unsw.edu.au, rasil.baidar@live.vu.edu.au, robert.abbas@vu.edu.au, sam.reisenfeld@mq.edu.au} \\
\textsuperscript{1}University of New South Wales, Sydney, Australia \\
\textsuperscript{2}Victoria University, Sydney, Australia \\
\textsuperscript{3}Macquarie University, Sydney, Australia
\end{tabular}
}
\maketitle

\begin{abstract}
The integration of Terrestrial Networks (TN) and Non-Terrestrial Networks (NTN), including 5G Advanced/6G-the Internet of Things (IoT), technologies,  using Low Earth orbit (LEO )satellites and high-altitude platforms (HAPS),  and Unmanned Aerial Vehicles (UAVs), is redefining the landscape of global connectivity. This paper introduces a new system-level security framework for 5G Advanced /6G, IoT Integrated Terrestrial Network-Non-Terrestrial Network (TN-NTN) with AI-native -Enabled Cloud Security. Due to the heterogeneity, scale, and distributed nature of these networks an new security challenges has been introduced to Leveraging AI native-driven cloud platforms offers powerful capabilities for real-time threat detection, security automation, and intelligent policy enforcement. The NTN satellite access function provides better security for discontinuous coverage via satellite connection. In addition to the above, the paper explores the security risks associated with the integrated 5G Advanced/6G IoT-TN-NTN with full network segmentation and network slicing systems and cloudification of RAN and Core, and presents a comprehensive AI-enabled cloud security framework.  We conclude with proposals for implementing AI-powered 5G Advanced/6G-IoT TN satellite-based NTN in future 5G Advanced/6G-IoT  networks and propose an approach that emphasizes zero-trust principles, federated learning, secure orchestration, a security framework, and resilience against adversarial threats.

.

\end{abstract}

\begin{IEEEkeywords}
5G Advanced, 6G, IoT,  Terrestrial Networks (TN),  Non-Terrestrial Networks (NTN), System Security, Cloud Security, AI Cloud Security, Network Slicing, LEO Satellite Services, Federated Learning, UAVs, Zero-Trust Architecture

\end{IEEEkeywords}

\section{Introduction}

The rapid advancement of mobile network technologies is driving the evolution of intelligent connected systems through the integration of advanced 5G / 6G networks, the Internet of Things (IoT) and non-terrestrial networks (NTNs), marking a pivotal leap forward. NTN's LEO satellite array provides global broadband connectivity, particularly in underserved locations. It uses a large number of low-Earth-orbit (LEO) satellites to provide high-speed internet with low latency. The system uses modern technology like as phased array antennas, optical inter-satellite communications, and digital processing to accomplish its performance. These integrated systems enable seamless connectivity between ground-based IoT devices and aerial or satellite-based infrastructure, creating the foundation for globally distributed cyber-physical systems.
Non-terrestrial networks (NTNs) have unique security challenges since they rely on distributed space-based infrastructure and integrate satellite and terrestrial networks. These concerns include the possibility of signal jamming, spoofing, and eavesdropping, as well as hardware and software flaws in the satellite. 
NTN (Non-Terrestrial Network) security refers to the safeguards used to ensure the integrity and confidentiality of data transferred over networks using satellites, high-altitude platforms (HAPS), and other non-terrestrial elements as part of 5G Advanced and future 6G infrastructure. These networks provide particular security difficulties due to their decentralized design and the physical properties of their components, such as satellites and UAVs.

While such integration enables applications in smart cities, remote healthcare, environmental monitoring, and autonomous vehicles, it also expands the system's vulnerability surface. Satellite IoT refers to using low-power, compact IoT devices that connect directly to satellites orbiting the Earth. There is no need for towers, cabling, or terrestrial infrastructure. These devices gather data from sensors on the ground, seas, or oceans and transmit it via satellite to the cloud for analysis. Traditional security approaches are inadequate to address the heterogeneity, scale, and dynamism of these systems. AI-powered cloud security offers the promise of adaptive, autonomous defense mechanisms capable of protecting vast, distributed infrastructures in real time.
This research paper will study the system security of  AI-powered TN and NTN  integrated systems.

\begin{itemize}
\item \textbf{Terrestrial Networks (TN) and Non-Terrestrial Networks (NTNs) }-Integrated 5G Advanced/6G-IoT NTN Systems 
Integrated architectures consist of :

\item \textbf{5G Advanced/6G RAN Networks}: consist of 5G Advanced UEs in flight or on the ground or the ships and 6G-capable service in the future with flying 5G  Advanced NBs,  with RUs and DUs for NTN, and traditional 5G Advanced NBs and 6G NBs for TN, where the path is from 5G Advanced (Releases 18-20) to 6G. The 3GPP standardization on 5G Advanced (Release 18-20) lays a solid foundation for 6G, including AI-native networks, integrated sensing and communications (ISAC), energy-efficient and sustainable architectures, energy-autonomous 6G IoT networks, and 6G systems with ubiquitous, three-dimensional connectivity.

Providing ultra-low latency, high bandwidth, and massive machine-type communications (mMTC).

\item \textbf {IoT Devices}: Including sensors, actuators, vehicles, and embedded systems with constrained resources.

\item \textbf {Non-Terrestrial Networks (NTNs)}: Utilizing mainly Low Earth Orbit (LEO), and a much smaller number of Medium Earth Orbit (MEO), and Geostationary Orbit (GEO) satellites, as well as High Altitude Platform Systems (HAPS).
NTN provides unique issues such as higher round-trip times, propagation delays, high Doppler shift, and frequent handovers.

\item \textbf {Cloud-Edge Infrastructure}: Managing orchestration, computation, and analytics through centralized or edge-deployed AI services.

This paper investigates the system-level security concerns of integrated 5G Advanced, 6G IoT and TN-NTNs and presents a comprehensive security framework leveraging AI-based cloud services.

\end{itemize}

\subsection{Background and Literature Review}

The rapid advancements in Non-Terrestrial Networks (NTNs) by moving from traditional Geostationary (GEO) satellites toward Low Earth Orbit
(LEO) satellites bring considerable improvements, specifically in terms of higher throughput, lower latency, and energy consumption [12]
 , particularly in the evolution towards 6G, have brought revolutionary connectivity solutions alongside significant cybersecurity concerns. Researchers have extensively explored these challenges, emphasizing innovative strategies to defend NTNs against sophisticated cyber threats, especially Distributed Denial-of-Service (DDoS) attacks. This review is structured into thematic subsections that summarize key advancements, highlight emerging trends, present recent innovations, and identify existing research gaps.
User terminals, especially mobile ones, can maintain excellent service quality and flexibly adjust to changing conditions by utilizing a variety of technologies, including free-space optical (FSO), radio frequency (RF), and hybrid FSO/RF linkages between dRAN units. This dynamic adaptability depends on a number of variables, such as FSO's reduction of air turbulence.
\begin{figure}
    \centering
    \includegraphics[width=1\linewidth]{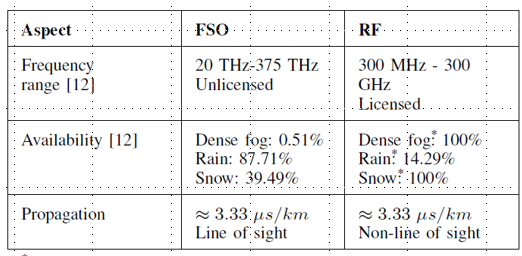}
    \caption{Comparison of FSO and communication links measured at 40 GHz}
    \label{fig:enter-label}
\end{figure}

\subsection{Anomaly Detection using AI}

Non-Terrestrial Networks (NTNs), which include satellites, unmanned aerial vehicles (UAVs), and high-altitude platforms (HAPS), represent a significant step toward providing the global, uninterrupted connectivity envisioned by 6G systems. Despite their obvious benefits, NTNs provide distinct and complex security challenges due to their inherent dynamic, dispersed, and resource-limited character. To ensure secure and dependable communication, the following major difficulties must be addressed:

\subsection{Cyberattack Vulnerabilities}
NTNs have inherent weaknesses, making them prime candidates for sophisticated cyber threats like as Distributed Denial of Service (DDoS) and man-in-the-middle (MitM) attacks. Given their widespread coverage and mobility, traditional security approaches are insufficient, leaving networks vulnerable to attackers who can disrupt crucial services like satellite internet or UAV-enabled emergency operations. For example, attacks against satellite networks such as Starlink could disrupt access in remote places with no terrestrial options. Similarly, MitM attacks take advantage of the wide NTN communication routes, which might compromise critical military conversations or secure financial transactions.

\subsection{Signal Interception and Jamming}
NTN systems predominantly rely on wireless transmissions, making them susceptible to interception (eavesdropping) and jamming threats. Unauthorized interception can severely breach data privacy, exacerbated by signal propagation characteristics over vast distances. Consequently, advanced encryption and signal protection methods become crucial. Jamming, an equally critical threat, deliberately disrupts signals such as GPS, affecting UAV navigation, potentially causing operational failures or mission disruption.

\subsection{Physical Infrastructure Threats}
Unlike ground-based networks, NTN infrastructure (satellites, drones, and ground stations) is physically accessible to attackers through both direct physical sabotage and remote cyber interference. Attacks targeting satellite ground stations or terminals can disable entire constellations, significantly impairing services. The remote and isolated nature of satellites and UAVs complicates traditional physical security measures. For example, targeted cyber-physical attacks on control stations could result in extended outages or severe operational degradation.

\subsection{Secure Data Transmission}
Securing data transmission is critical, given the sensitive information carried by NTNs, including emergency services, governmental communications, and IoT data. Dynamic NTN topologies complicate encryption, authentication, and key management. Traditional encryption approaches may introduce latency and excessive computational overhead, unsuitable for rapid NTN communications. Therefore, research into lightweight encryption, blockchain-based secure frameworks, and quantum cryptography-driven key management schemes is essential.

\subsection{Real-Time Security Adaptation}
Real-time security adaptation is especially challenging in NTNs due to their rapidly changing network structures. Static security solutions quickly become ineffective, necessitating fast threat identification and dynamic response mechanisms. Delayed responses exacerbate threats, potentially causing widespread disruption. Thus, deploying AI-driven and deep learning-based anomaly detection tools is crucial, enabling rapid, dynamic security adjustments to emerging threats.

\subsection{Resource Constraints and Computational Limitations}
NTN components, particularly satellites and UAVs, are constrained by limited computational resources, bandwidth, and power. Implementing complex security technologies, such as deep learning models or blockchain, can be burdensome. Addressing this challenge requires developing efficient, lightweight, and computationally feasible security frameworks specifically designed for NTNs. Technologies like edge computing and decentralized architectures are promising, reducing computational burdens and latency.

\subsection{Dynamic Topology and Network Management Complexity}
The constantly evolving NTN topology, driven by satellite movements, UAV trajectories, and dynamic user interactions, complicates security management. Unlike fixed terrestrial networks, NTNs must continually adapt their security posture to real-time changes, complicating network monitoring, threat detection, and incident response. Advanced AI techniques, including self-organizing maps (SOM), reinforcement learning (RL), and transformer-based attention models, offer the required adaptability to maintain robust security amidst dynamic NTN configurations.

\subsection{AI in TN-NTN Security}
Machine learning (ML) and deep learning (DL) are increasingly essential in cybersecurity for dynamic NTNs, where traditional methods fall short. Prominent contributions include
\begin{itemize}
\item Dhaliwal et al. (2018) demonstrated XGBoost's effectiveness in accurately and efficiently detecting network intrusions.
\item Lam and Abbas (2020) introduced ML-based anomaly detection specifically designed for 5G Advanced networks, providing foundations relevant to NTNs.
\item Amjad Iqbal et al. (2024)
Threat Detection and Prevention: Artificial intelligence can be used to detect and prevent network threats in real time, reducing the risk of attacks. Anomaly detection: Artificial intelligence can be used to detect and report anomalous network behavior that may indicate a security vulnerability. Predictive analysis: Artificial intelligence can be used to identify potential security concerns, enabling preventive actions to be taken before an attack occurs. Predictive analysis: Artificial intelligence can be used to identify potential security concerns, enabling preventive actions to be taken before an attack occurs [23].

\item Shrestha et al. (2021) developed ML-powered intrusion detection for UAV cellular networks, significantly enhancing detection speed and accuracy.
\item Pokhrel et al. (2021) utilized ML techniques to detect IoT botnets, pertinent to NTN contexts like smart agriculture and urban scenarios.
\item Tariq et al. (2020) employed Random Forest algorithms for rapid classification of attacks, showcasing efficiency within NTN cybersecurity frameworks.
\end{itemize}

\subsection{Emerging Trends: Quantum Cryptography, Blockchain, and Edge Computing}
Advanced cryptographic methods and AI-based security solutions are enhancing NTN security substantially:
\begin{itemize}
\item Ylianttila et al. (2020) emphasized Quantum Key Distribution (QKD) for secure data transmissions, addressing interception threats effectively.
\item Shakya et al. (2025) proposed a zero-touch, zero-trust AI framework, reducing manual interventions and improving response times, critical in dynamic NTNs.
\item Dibaei et al. (2020) analyzed vulnerabilities in intelligent connected vehicles (ICVs), parallel to UAV-based NTN applications, highlighting integrated cyber-physical protection strategies.
\item Kumar et al. (2022) demonstrated blockchain-based decentralized solutions to authenticate and secure NTN nodes effectively.
\item Wang et al. (2022) explored edge computing to alleviate computational latency and resource limitations inherent to NTN environments.
\end{itemize}

\subsection{Recent Innovations and Applications}
Recent literature highlights the growing reliance on sophisticated AI techniques:
\begin{itemize}
\item Ali et al. (2020) forecasted deep learning's pivotal role in automated threat detection, network optimization, and predictive security measures within 6G NTNs.
\item Jiang et al. (2023) showcased transformer-based models adept at identifying subtle, sequential, and low-rate attack patterns typical in advanced persistent threats against NTNs.

\item To improve connection reliability, we use a combination of Free-Space Optical (FSO) and Radio Frequency (RF) communication methods for linky [20].

\end{itemize}

Key insights from recent studies include the growing importance of automation, predictive analytics, and transformer models for sophisticated threat detection.

\subsection{Limitations and Research Gaps}
Despite considerable progress, several critical gaps remain:
\begin{itemize}
\item Mahboob and Liu (2024) pointed out the shortage of NTN-specific models that effectively address unique challenges like high latency, mobility, and resource constraints.
\item Singh et al. (2023) noted that existing datasets often lack realism concerning satellite and UAV communication scenarios, advocating for specialized synthetic datasets.
\item Ahmad et al. (2022) highlighted a deficiency in ML frameworks integrating hybrid methodologies capable of simultaneously detecting high-volume and subtle, low-rate attacks.
\end{itemize}

\begin{figure}[htbp]
\centerline{\includegraphics[width=\linewidth]{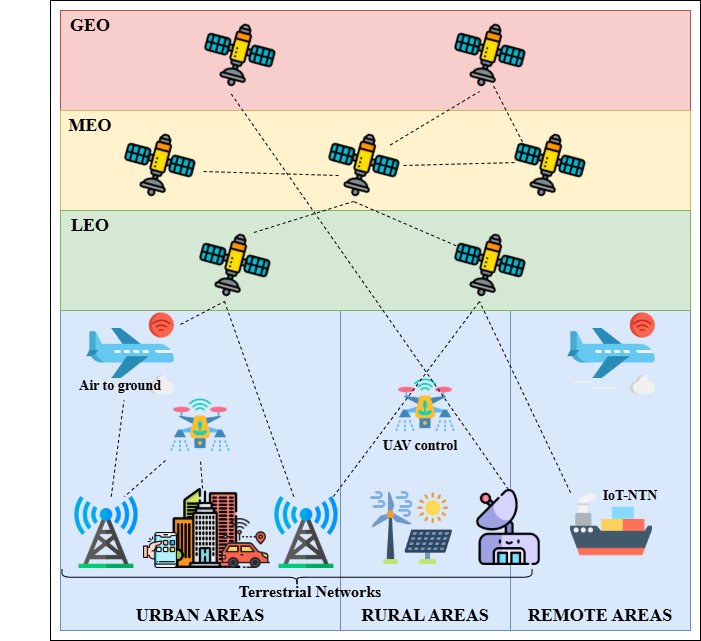}}
\caption{NTN network}
\label{fig:heatmap}
\end{figure}

Fig.1 presents a comprehensive architecture of a Space–Air–Ground Integrated Network (SAGIN), designed to deliver seamless, resilient, and low-latency connectivity through the integration of Non-Terrestrial Networks (NTNs) with terrestrial communication infrastructure. This multi-layered architecture spans geostationary (GEO), medium Earth orbit (MEO), and low Earth orbit (LEO) satellite constellations, aerial platforms including unmanned aerial vehicles (UAVs), and ground-based terrestrial networks. The system is engineered to support diverse communication requirements across urban, rural, and remote regions, ensuring ubiquitous access and service continuity regardless of location or infrastructure availability.    
\begin{figure}
    \centering
    \includegraphics[width=1\linewidth]{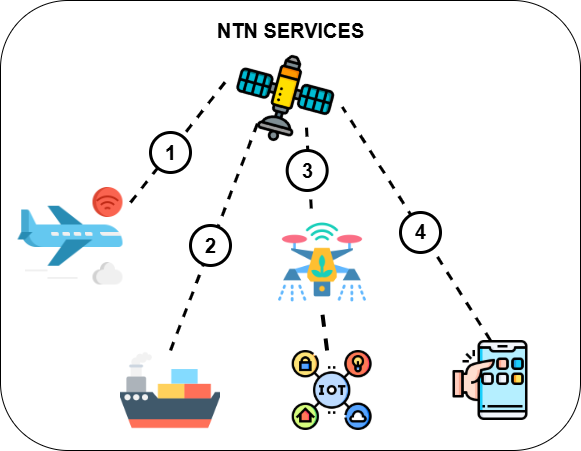}
    \caption{NTN services}
    \label{fig:enter-label}
\end{figure}
The space segment includes satellites at three orbital layers, each providing different capabilities in terms of coverage, latency, and capacity. GEO satellites, positioned approximately 35,786 kilometers above the equator, offer extensive coverage and are commonly used for broadcasting and high-latency services such as global backhaul. Despite their wide footprint, the significant propagation delay (~600 milliseconds round-trip) limits their use in latency-sensitive applications. MEO satellites, typically operating between 2,000 and 20,000 kilometers in altitude, offer a balance between latency and coverage and are often employed in navigation and moderate-latency services. LEO satellites, located at altitudes below 2,000 kilometers, provide low-latency (20–40 milliseconds) and high-throughput connectivity due to their proximity to the Earth. These satellites are particularly suitable for real-time communications, broadband internet delivery, and Internet of Things (IoT) applications in remote areas. Inter-satellite links (ISLs) enable dynamic routing and load balancing across these layers, forming a resilient mesh network in space.

The airborne segment is composed of both manned and unmanned aerial platforms that act as agile communication relays and data collection nodes. Air-to-ground (A2G) communication supports high-speed connectivity for in-flight aircraft, enabling real-time data transmission and access to network services during flight. UAVs function as edge nodes and relay agents, particularly in scenarios where terrestrial infrastructure is absent or damaged. They play a vital role in supporting rural and remote communications by forming airborne ad hoc networks that link to LEO satellites or nearby ground stations, thereby extending network reach and enhancing operational flexibility.

On the ground, the terrestrial segment is organized by geographic context into urban, rural, and remote regions. Urban areas benefit from dense network infrastructure, including macro and small cell base stations, roadside units, and intelligent IoT deployments. Here, NTNs can serve as an offloading mechanism or redundancy layer, especially during peak demand or in disaster scenarios. Rural areas, often underserved by conventional infrastructure, rely on a hybrid approach that integrates UAV relays, solar-powered base stations, and satellite backhaul to maintain connectivity. In these areas, coordinated satellite-UAV-ground schemes are essential for ensuring service quality and reliability. Remote areas, which typically lack any terrestrial infrastructure, are dependent on NTN components such as LEO constellations and UAVs for communication. This is particularly critical for applications like maritime tracking, environmental monitoring, and mission-critical IoT systems, where connectivity must be ensured despite geographic isolation.

The integrated SAGIN-NTN framework supports a broad spectrum of use cases and service-level requirements. These include low-latency control for UAV missions, massive machine-type communication (mMTC) for IoT sensor networks, enhanced mobile broadband (eMBB) for underserved areas, and ultra-reliable low-latency communication (URLLC) for time-sensitive and safety-critical applications such as emergency response and surveillance. The architecture is designed to support dynamic network reconfiguration, adaptive routing, and efficient spectrum utilization across layers and domains.

\section{System Architecture}
\begin{figure}
    \centering
    \includegraphics[width=1\linewidth]{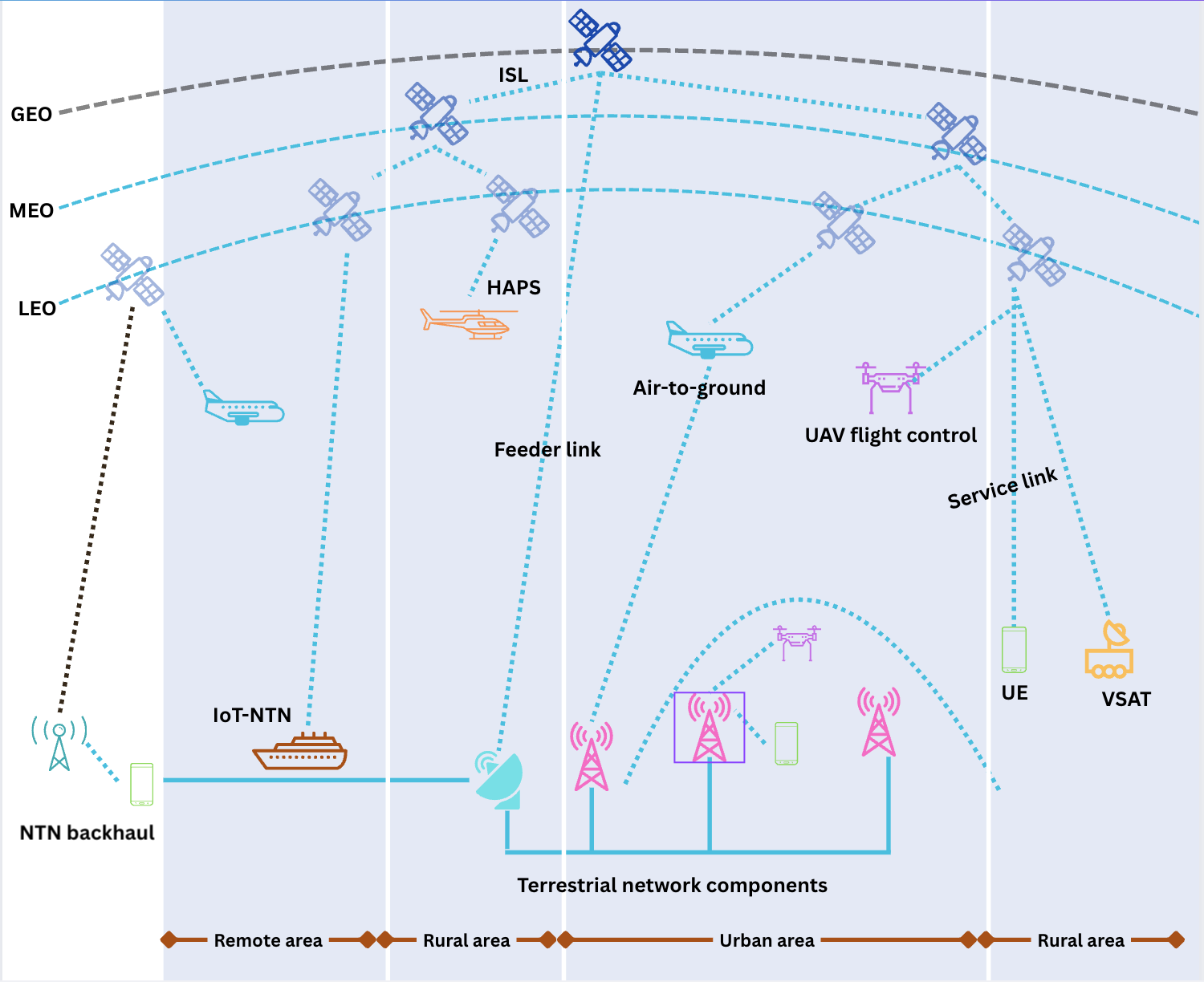}
    \caption{5G Advanced/6G/IoT TN-NTN services}
    \label{fig:enter-label}
\end{figure}enables the connection of 5G Advanced devices to satellites, which is particularly beneficial for disaster recovery or in remote locations.  LEO satellites and 5G Advanced/6G flying base stations are expected to be implemented alongside terrestrial cell towers, 5G Advanced Non-Terrestrial Networks (NTNs) use satellites and other space-based platforms. 5G Advanced NTN was formally standardized by 3GPP Release 17, which also supported NB-IoT/eMTC and 5G Advanced New Radio (NR). This makes it possible to link 5G Advanced Advanced devices to satellites, which is very helpful for disaster recovery or in remote locations, and it is expected that LEO satellites and 5G Advanced Advanced/6G flying base stations will be implemented in 2030. 

Integrating disaggregated RAN (dRAN) with Non-Terrestrial Networks (NTNs), including satellites, is crucial for seamless 6G communication with advanced terrestrial 5G networks. Low Earth orbit (LEO) satellites function in a circular path around Earth, positioned at altitudes between approximately 321 km and 1500 km, moving at speeds of about 28,000 km/h, and completing an orbit in roughly 90 minutes.  
Because their altitude is relatively low, launching LEO satellites does not require powerful rockets, which makes the process more economically viable.  
Another clear benefit of the shorter distance is the reduced round-trip time (RTT), usually less than 30 ms.  
LEO satellites are generally compact, often measuring less than 1 meter in perimeter or even down to a few centimeters for nano-satellites, with weights below 500 kg.  
It is assumed that the NTN employs a beamforming technique at the satellite station, with a typical beam footprint ranging from 100 km to 1000 km.  
\begin{figure}
    \centering
    \includegraphics[width=1\linewidth]{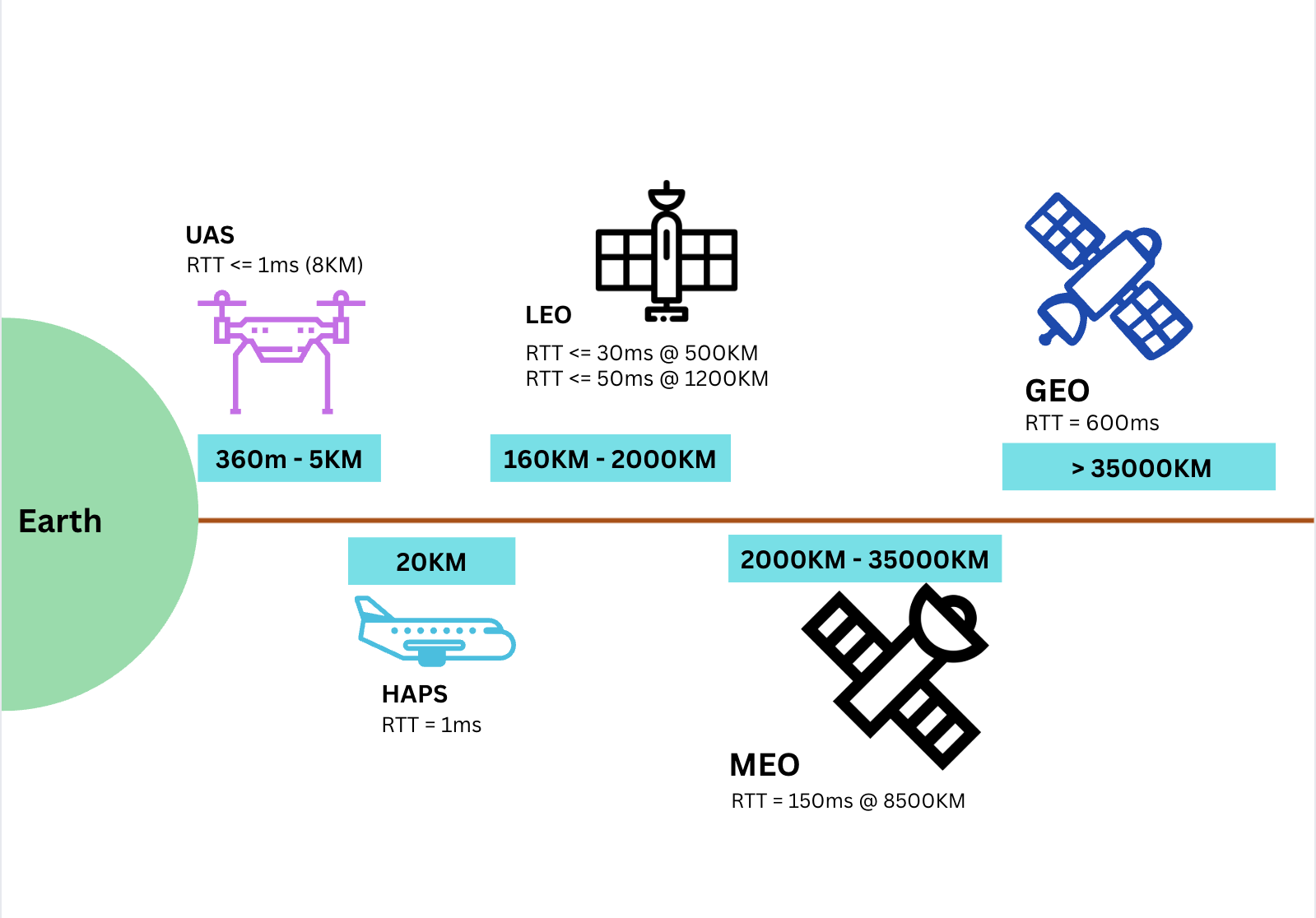}
    \caption{RTT for NTN System}
    \label{fig:enter-label}
\end{figure}
One disadvantage of LEO satellites is their operation in challenging environments, as atmospheric gases may still be present, leading to an estimated lifespan of under 10 years.

\subsection{5G Advanced, 6G/IoT-TN and service use cases}

The 5G Advanced/IoT-TN has been implemented recently, and 5G has been in use since 2020, utilizing 3GPP Release 15. The 6G standardization work began in 2024 with Release 19 and is ongoing, focusing on requirements, and is expected to be implemented in 2030.
To fully access 5G Advanced features, a 5G  Advanced Standalone (SA) core is required, as several essential features of Release 18 rely on capabilities only available in a 5G Advanced SA core.
A 5G Advanced Standalone (SA) Core is required for full access to 5G Advanced functionalities. Many essential aspects of Release 18 rely on capabilities that are only available with a 5G Advanced Core. It is necessary for intelligent network slicing, URLLC, AI/ML-based network optimization, and advanced QoS handling. 

The 5G Advanced SA Core's modular, cloud-native service-based architecture allows for dynamic service composition and exposure, resulting in these enhanced features.

\subsection{5G Advanced RAN and 6G Core}

To facilitate a seamless transition to 6G and prevent market fragmentation, the specifications for 6G radio access technology (RAT) should be defined exclusively in standalone mode, with user equipment connecting solely to 6G from the outset. Additional important design principles for the future 6G RAN entail creating open interfaces between the RAN and other network domains to foster a robust ecosystem, adopting an AI-native approach to ensure seamless application of AI/ML whenever suitable, enhancing energy efficiency in the 5G Advanced Advanced RAN, and providing improved support for crucial verticals and deployment scenarios right from the beginning, including non-terrestrial access, Massive IoT, and time-sensitive communication services. Cloud-native 5G/6G applications and use cases are specifically tuned for cloud-native infrastructure. When compared to operating containers in virtual machines, the infrastructure solution offers a drastically simplified architecture by enabling the deployment of containers as a service (CaaS) directly over bare metal servers, eliminating the need for a virtualization layer.
\begin{figure*}[h!]
    \centering
    \includegraphics[width=0.7\linewidth]{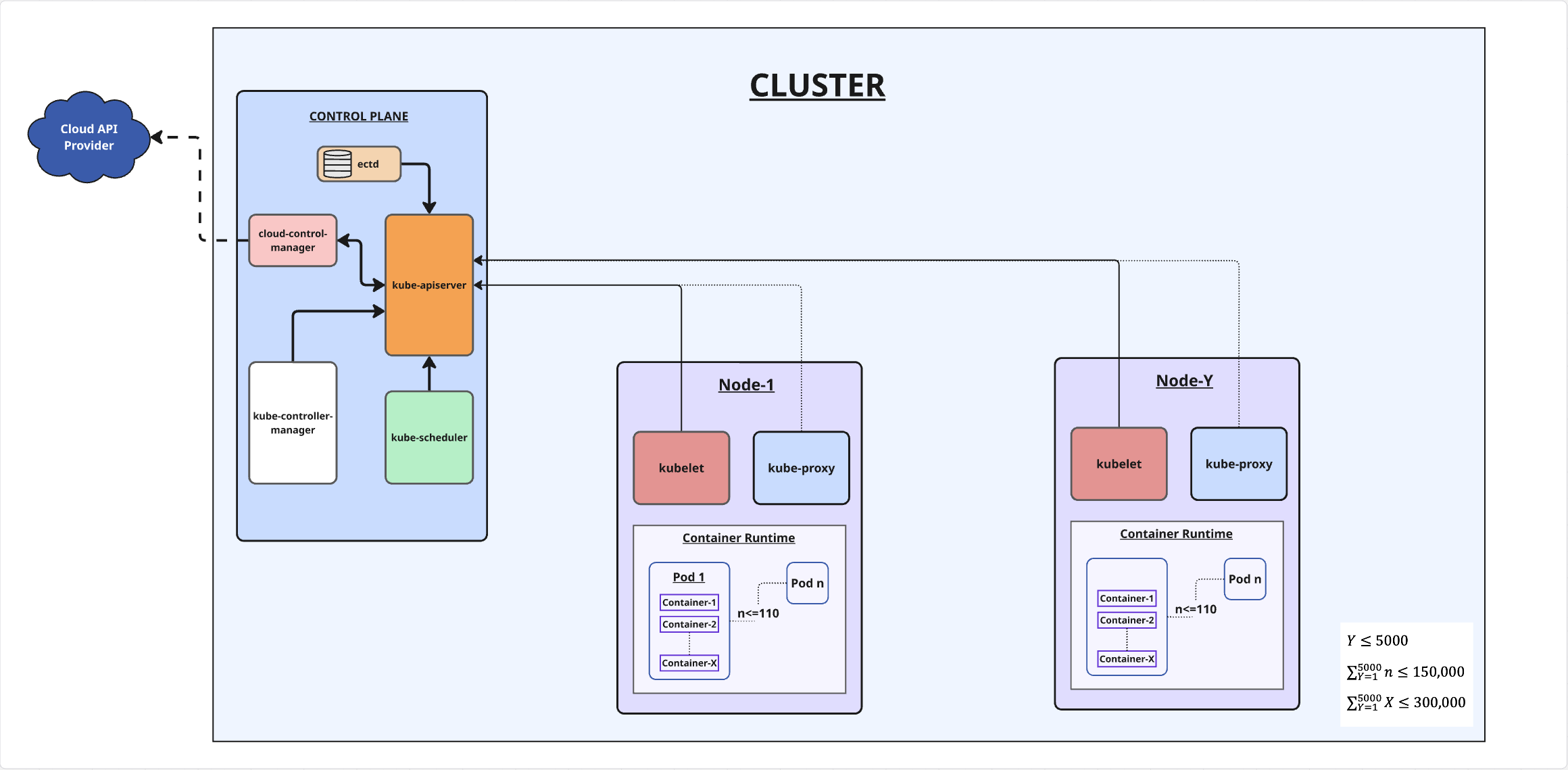}
    \caption{Concepts of a Kubernetes architecture
    }
    \label{fig:kubernetes-architecture}
\end{figure*}
According to Kubernetes v1.33, a Kubernetes cluster architecture (see Fig.~\ref{fig:kubernetes-architecture}), each cluster is limited to the following cluster-wide scale constraints~\cite{KubernetesIO}:-

\begin{itemize}
  \item Nodes ($Y$): $Y \le 5\,000$
  \item Pods ($n$): $\displaystyle \sum_{i=1}^{Y} n_{i} \le 150\,000$
  \item Containers ($X$): $\displaystyle \sum_{\text{all Pods}} X_{i} \le 300\,000$
\end{itemize}

The 6G Core will be an advancement of the current 5G Advanced Core framework, serving as a fundamental component of the future 6G network design. This will ensure that essential 5G Advanced Core network functionalities, such as exposure, network slicing, intergenerational interoperability, and roaming, are available from the very beginning of the 6G deployment. Furthermore, the core service-based architecture is expected to undergo further development aimed at reducing system-level complexity when integrating new network functions that are applicable to both 5G Advanced and 6G, as well as those exclusive to 6G.

\begin{figure}
    \centering
    \includegraphics[width=1\linewidth]{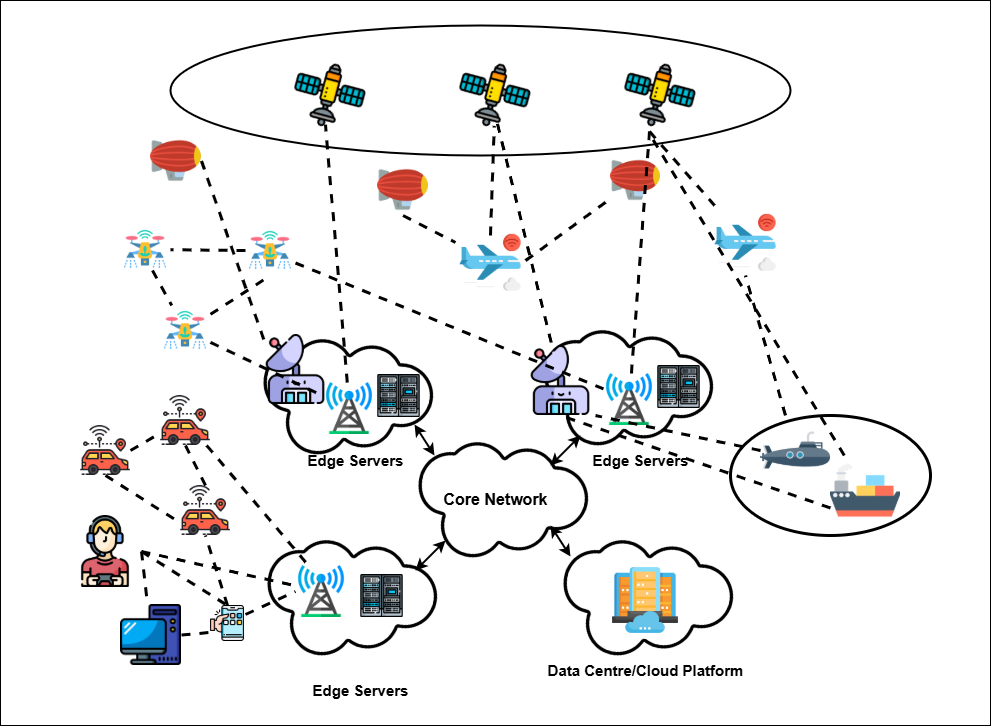}
    \caption{5G Advanced/6G/IoT Cloud architecture
    }
    \label{fig:enter-label}
\end{figure}

\section{System Security}

5G Advanced/6G/IoT security: evolving to meet future network challenges
High-trust cyber-physical systems that link humans with intelligent machines demand exceptional reliability and resilience, accurate positioning and sensing, as well as ultra-low-latency communication. This imposes significant requirements on the security capabilities of 5G Advanced/6G/IoT TN-NTN systems, as well as on its performance, including the capacity to ensure that the necessary features are implemented.

\begin{figure}
architecture.jpg    \centering
    \includegraphics[width=1\linewidth]{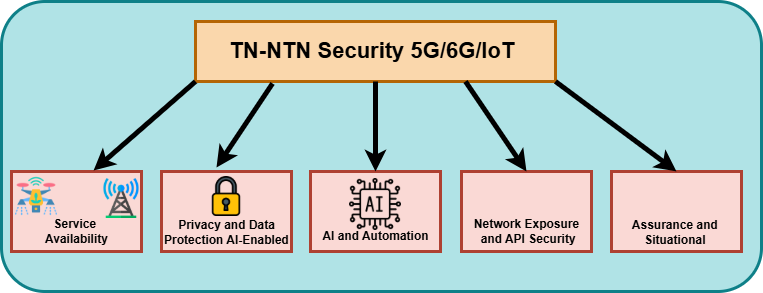}
    \caption{NT-TNT System Security Components}
    \label{fig:enter-label}
\end{figure}

6G networks must provide this assurance to both users and service providers through enhanced awareness and security resilience in both the deployment and operational stages. On a personal level, the security features of 6G must honor privacy and the ownership of personal data in an interconnected world. These capabilities should be robust, yet easily customizable to suit individual user preferences.

Mobile networks are becoming an increasingly important aspect of society, necessitating stricter security and availability standards. As the cyber and physical worlds converge and the use of mobile networks evolves beyond communication, 6G security will need to adopt a more holistic approach, taking into account both communication and computing more than previous generations. Building on the security of 5G Advanced, the security of the 6G network will be built on open standards, with a greater emphasis on operational issues. Furthermore, new use cases and technologies, such as immersive communication and zero-energy gadgets, will need updated threat assessments and new security solutions.
In the TN-NTN era, it's crucial to prioritize system-wide security, including mobile networks, by knowing the following:

• Higher stakes and lower risk tolerance 
• Dependence on both cyber and physical systems
• Ensure security for standards, goods, deployments, and operations. 
• Implement proactive cybersecurity measures. 
• Manage vulnerabilities. 
• Secure the supply chain.

Implementing advanced-Advanced 5G requires strong network security, as major enablers such as network slicing, edge computing, AI-driven automation, and API-based capability exposure extend the threat surface and increase the risk of cyber attacks. 

5G Advanced requires strong identity management, slice isolation, edge-level data protection, and safe orchestration of virtualized network operations in dynamic, programmable systems with multiple tenants. Enterprise and mission-critical use cases require rigorous security assurances to meet regulatory standards and build confidence across industries. 

Evolving from 5G Advanced TN security, the security for 6G TN and NTN will use additional measures to respond to an evolving threat landscape and to support new use cases, shifts in technology and society, and regulatory requirements. With mobile networks increasingly being considered as part of critical infrastructure and with the cyber-physical merge emphasizing the role of connectivity in everyday life, 6G security will need to be based on a holistic view, combining different solutions and approaches to gain situational awareness. Automated security and threat management, operational assurance, and mechanisms that adhere to zero-trust principles will contribute to situational awareness.

\begin{figure*}[h!]
\centering
\includegraphics[width=0.8\textwidth]{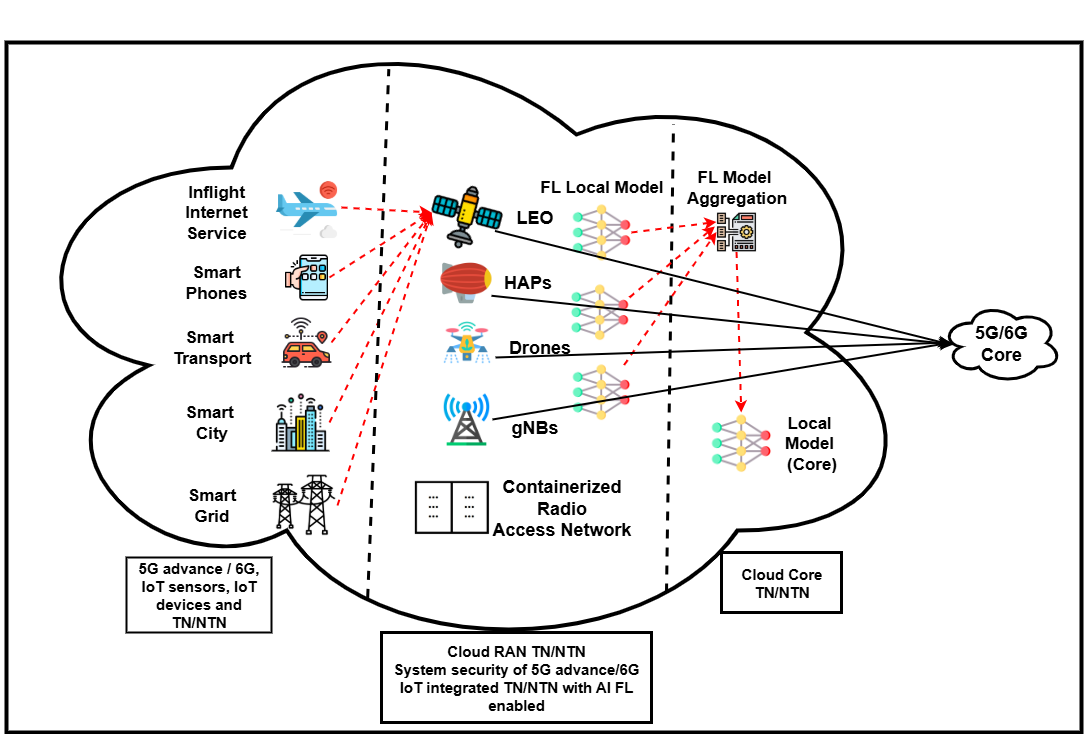}
\caption{System security of 5G Advanced /6GIoT integrated TN-NTNwith AI FL enabled cloud security}
\label{fig:model_comparison}
\end{figure*}

The virtualization and automation of network functions have facilitated the deployment of mobile cores in the cloud while also opening up avenues for the cloud-based implementation of open Radio Access Networks (RAN). This paper investigates the threats, vulnerabilities, and security measures associated with Cloud RAN implementations. Alongside conventional attacks on the RAN and Core, the cloud infrastructure's vulnerabilities, including microservices, container engines, host operating systems (OS), and third-party hardware, may be exploited in cloud-based RAN and Core scenarios. Attack vectors like supply chain attacks, cross-container breaches, insufficient authentication, and misconfigurations can be leveraged by both internal and external threat actors to compromise applications and access data during transit, in use, and at rest, leading to attacks on confidentiality, integrity, and availability. Botnets, as well as distributed volumetric denial-of-service (DDoS) and application DDoS attacks targeting the user plane from both internal devices and external attackers, can affect the availability of cloud services and cloud-based applications [22].

 The Kubernetes architecture is a sophisticated orchestration of components that work together to provide a resilient, scalable, and manageable environment for modern applications. Its design encapsulates the complexity of distributed data. allowing DevOps teams to focus on application-level concerns rather than the intricacies of the underlying infrastructure. Kubernetes is Cloud Native; VMs are not.
 Cloud-native applications benefit from performance containers that are faster, lighter, and more efficient. Containers have a startup time of milliseconds to seconds, while VMs take tens of seconds to minutes. In addition, containers offer efficient resource utilization with no OS overhead.

\subsection{ Network segmentation and Network Slicing}
To improve cybersecurity performance , network segmentation and network slicing are used.
Network segmentation refers to the process of splitting a larger network into smaller, isolated pieces or subnets. This method tries to mitigate the impact of security breaches by isolating possible threats inside certain segments and preventing them from spreading throughout the network.

Standalone network slicing, an essential capability of 5G Advanced Standalone (SA) networks, enables network providers to establish various virtual networks (slices) within a single physical network framework. Each slice can be configured autonomously with tailored QoS requirements, facilitating personalized experiences for various services and applications. This approach optimizes the use of network resources and allows for customized connectivity to meet different demands, such as low-latency communications or high-bandwidth video streaming, and improves security through slice-and-use case applications separation and isolation.
End-to-end network slicing, a key feature of 5G Advanced and 6G, focuses on constructing segregated, virtualized networks over a shared physical infrastructure. Each "slice" is designed to meet the requirements of a certain application or customer.
When combined with cloud security, this feature provides significant benefits that extend far beyond performance—it fundamentally improves security posture.

\begin{figure*}[h!]
\centering
\includegraphics[width=1\textwidth]{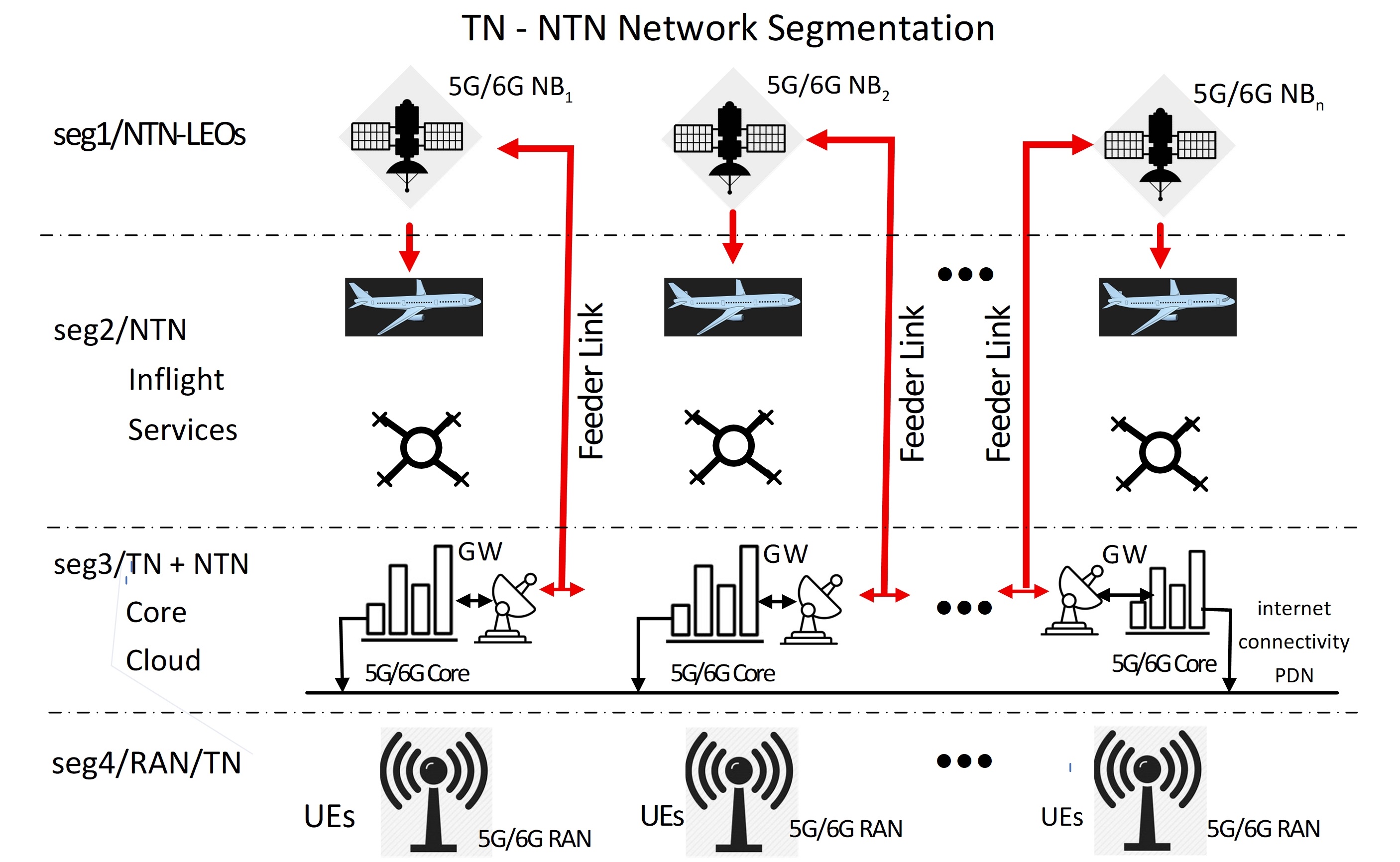}
\caption{TN-NTN Network Segmentation}
\label{fig:model_comparison}
\end{figure*}

\subsection{Cloud Security (TN-NTN)}
As technology advances, adversaries become more savvy, exploiting vulnerabilities in today's fragmented security landscape 5G advanced /6G/IoT networks remain reliable and secure 5G Advanced whenever possible. However, there are significant security and privacy challenges due to the wide range and quantity of devices and open network nodes. In order to address these issues, artificial intelligence and machine learning will not only bring about novel methods for network orchestration and service management, but they will also safeguard network endpoints and security gateways.
\begin{figure}
    \centering
    \includegraphics[width=0.8\linewidth]{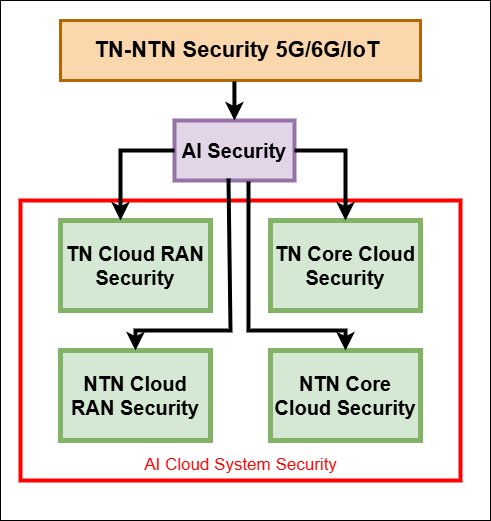}
    \caption{5G Advanced Advanced/6G/IoT Security Cloudification }
    \label{fig:enter-label}
\end{figure}
A TN-NTN cloud environment is just as safe as its weakest link. Thus, successful cloud security requires many technologies to work together to protect data and applications from all sides. This frequently comprises firewalls, identity and access management (IAM), segmentation, and encryption.

Rather than protecting a perimeter, cloud security safeguards resources and data separately. This entails installing more detailed security measures, such as cloud security posture management (CSPM), data protection, data security, disaster recovery, and compliance software.

Cloud infrastructures, particularly hybrid clouds that include public clouds and private data centers, might have numerous internal and external risks. To make them accessible and secure, it is necessary to use access restrictions, multifactor authentication, data protection, encryption, configuration management, and other technologies.
The advanced deep packet inspection solutions are based on machine and deep learning (ML/DL) algorithms in conjunction with behavioral, statistical, and heuristic techniques. Even in encrypted, obfuscated, or anonymized scenarios, both solutions provide real-time, thorough insights into the protocol, application, and service type for every network flow. With vector-ready APIs and CPU improvements, it is especially designed for cloud computing environments.
Mobile Edge Computing (MEC) was developed in response to the growing need for low-latency and energy-efficient mobile applications. MEC allows computing tasks to be transferred from resource-constrained mobile devices (MDs) to adjacent edge servers, which is suitable for NTN and smart transport use cases. 
To improve system performance, the offloading for Multi-User MEC (Multi-access Edge Computing) is used to involve many users (devices) offloading computationally expensive work from their local devices (e.g., smartphones, IoT sensors, smart transport sensors, and AR/VR devices) to edge servers located closer to the user than centralized cloud data centers. This reduces latency, increases performance, and saves energy on the device.

MEC extends cloud computing capabilities to the network's edge, near base stations and access points.

It enables low-latency, high-bandwidth services for TN-NTN 5G/6G IoT, AR/VR, and AI-powered applications. Use case examples:

\begin{itemize}
    \item {\textbf{AR/VR streaming}} for multiple users in smart stadiums or events.
    \item \textbf{In-flight internet services}: NTN-Broadband Internet
    \item \textbf{Autonomous Vehicles,} where multiple vehicles share road condition data at the edge.
    \item \textbf{Industrial IoT} with robotic coordination via edge servers.
    \item \textbf{Smart Cities} for real-time video analytics from multiple surveillance cameras.
    \item \textbf{Smart Shipping} IoT/ NTN
\end{itemize}

\subsection{ AI in TN-NTN Cloud Security}

\textbf{AI-powered cloud security} refers to the integration of artificial intelligence (AI) and machine learning (ML) technologies to protect cloud-based systems, data, and infrastructure from cyber threats. It enhances traditional cloud security by making it more \textbf{adaptive, intelligent, and automated}.

AI is already transforming network management, optimization, and customer experience algorithms. AI is integrated and standardized in 5G Advanced, setting the groundwork for 6G. 
\begin{figure}[!ht]
    \centering
    \includegraphics[width=0.9\linewidth]{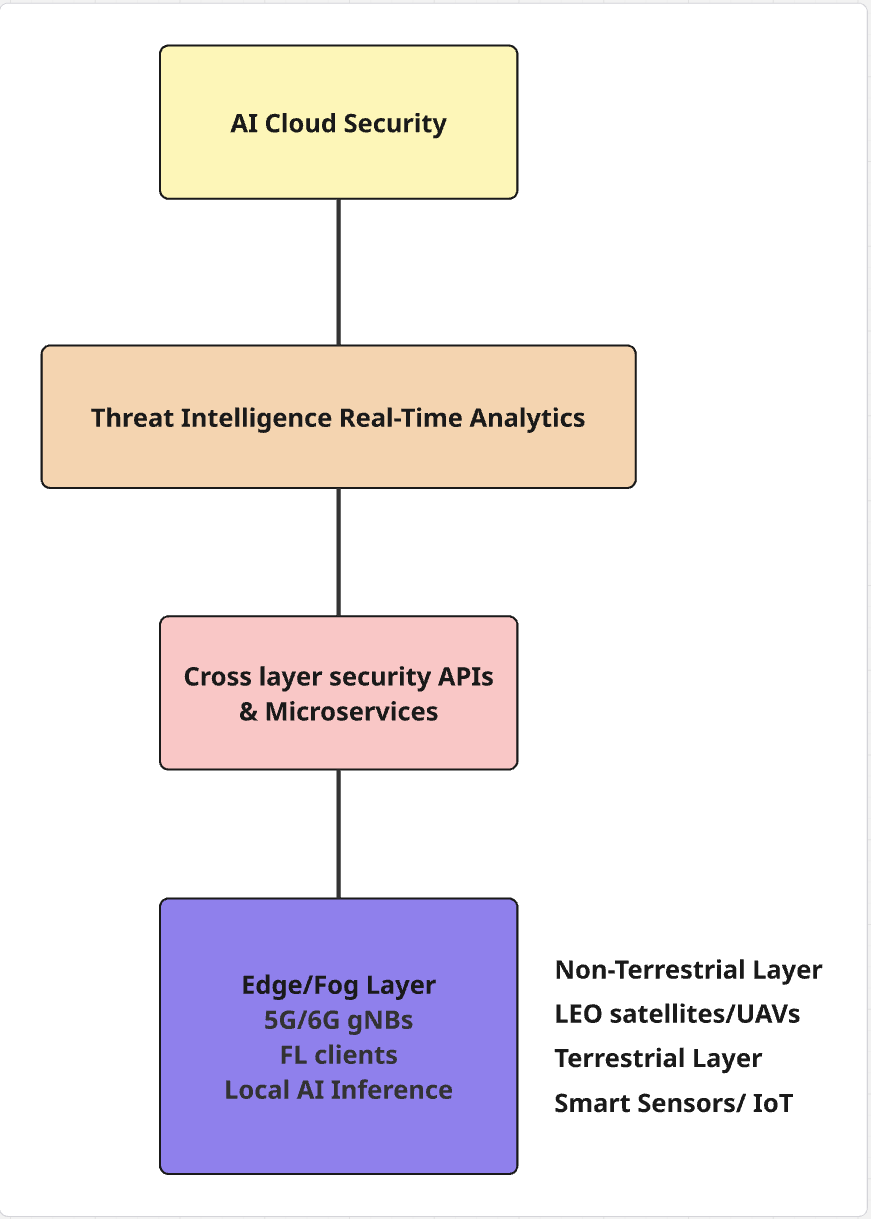}
    \caption{Integrated TN-NTN-FML Architecture.jpg}
    \label{fig:ai-cloud-security-framework-tn-ntn-iot}
\end{figure}
 Network security is critical for adopting 5G Advanced/6G since fundamental enablers such as network slicing, edge computing, AI-driven automation, and API-based capability greatly broaden the danger surface and increase the risk of cyberattacks.
5G Advanced Advanced-Advanced requires strong identity management, slice isolation, edge-level data protection, and safe orchestration of virtualized network operations in dynamic programmable systems with multiple tenants.
 Enterprise and mission-critical use cases require rigorous security assurances to meet regulatory standards and build confidence across industries.

\section{AI-Enabled Cloud Security Framework}

\subsection{Zero Trust Architecture}

A feature of zero trust analyzes 3GPP 5G Advanced /6G security scenarios for the 5G Advanced/6G core network that could benefit from the zero trust principle and identifies associated dangers.
Zero-trust should be enforced in important network slices and edge services. Regulations governing cross-border transactions must prioritize security and privacy. Data are trained to ensure AI/ML-based decisions are visible, auditable, and traceable. This covers logging events, data sources, and intent signals. Mandatory AI/ML model certification is necessary, together with legal frameworks to control model ownership and prevent fraud. 
\begin{figure*}[h!]
    \centering
    \includegraphics[width=0.7\linewidth]{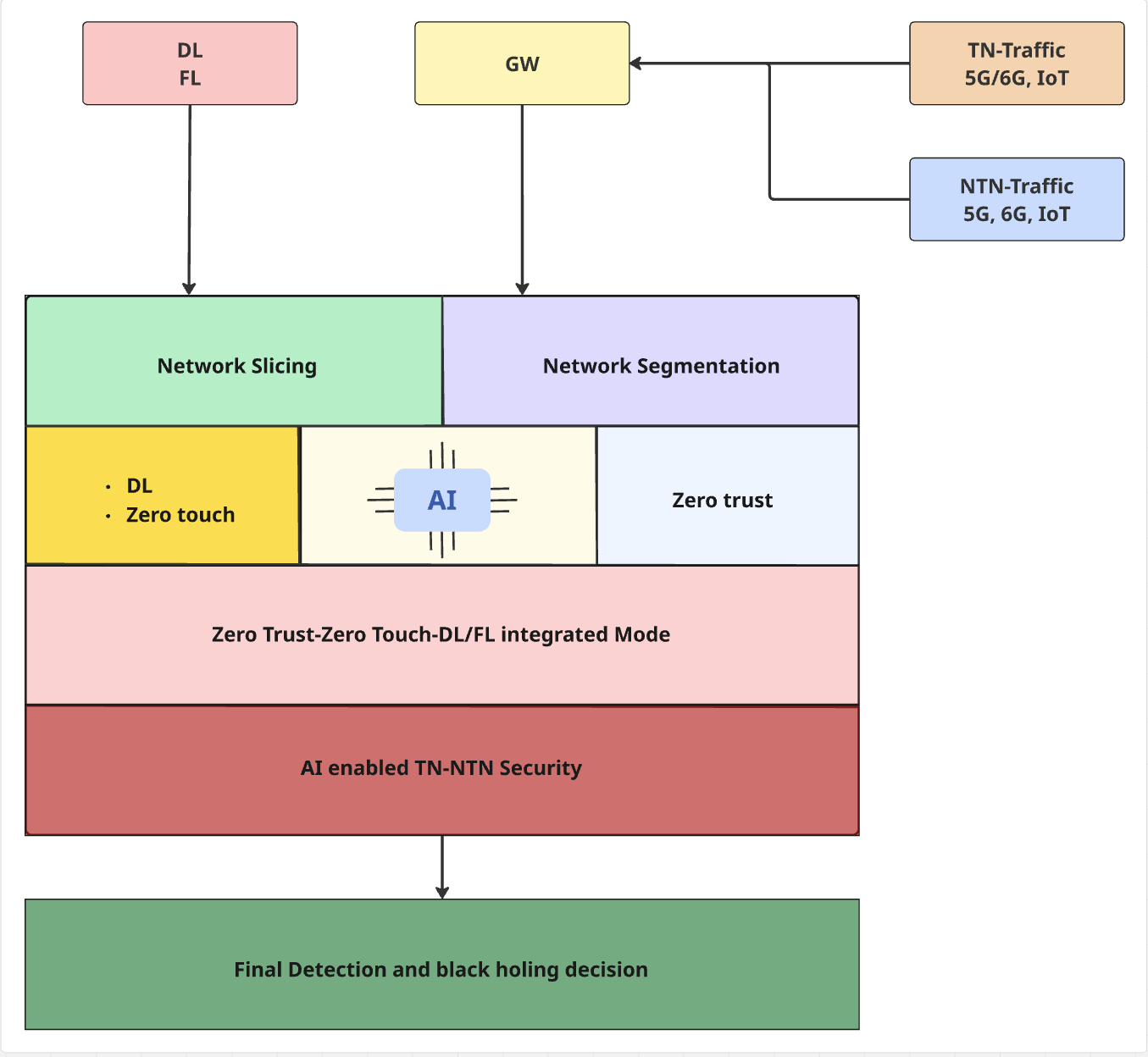}
    \caption{AI-enabled TN-NTN  Automated Security Framework}
    \label{fig:placeholder}
\end{figure*}
\subsection{Federated Learning (FL) in Non-Terrestrial Networks}
Federated learning is a decentralized, privacy-preserving machine learning technique that enables multiple devices or nodes (such as base stations borne on satellites, edge servers, mobile devices, and Internet of Things devices in flights and on the ground remote areas or ocean ships) to collaborate in training a common model without exchanging raw data. Mobile devices - UEs with limited resources to adjacent edge servers.

(NTNs) involves training machine learning models across distant devices or servers in NTNs without sharing raw data, hence increasing privacy and enabling collaborative learning. This method uses NTNs, including satellite constellations and aerial platforms, to promote communication and training between various devices and servers. Using FL in NTNs preserves data privacy and allows for more effective training on varied, geographically scattered datasets. 
Federated Learning (FL) in Non-Terrestrial Networks (NTNs) entails training machine learning models across distant devices or servers in NTNs without sharing raw data, hence increasing privacy and enabling collaborative learning. This method uses NTNs, including satellite constellations and aerial platforms, to promote communication and training between various devices and servers. Using FL in NTNs preserves data privacy and allows for more effective training on varied, geographically scattered datasets.

\subsection{Federated Learning for Edge}
A distributed machine learning technique called federated learning uses several servers or devices to work together to build a model without disclosing their raw data. This makes it possible to train models across decentralized datasets while protecting privacy, facilitating cooperation, and upholding data sovereignty. Edge devices (gNBs/satellites) train lightweight models (for
example, TinyML-LSTM) on local IoT data streams.
The merging of Terrestrial Networks (TN) and Non-Terrestrial Networks (NTN) into 5G Advanced Advanced and 6G systems provides worldwide coverage and seamless connectivity for IoT applications. However, the huge scale, distributed nature, and sensitivity of IoT data present substantial obstacles to centralized cloud-based machine learning. Federated Learning (FL) is emerging as a promising paradigm for cooperatively training machine learning models without sharing raw data. This section examines the use of FL in TN-NTN edge contexts, emphasizing its role in improving privacy, lowering latency, and optimizing bandwidth utilization.

transforms TN-NTN 5G Advanced Advanced/6G IoT by enabling \textbf{secure, low-latency intelligence at the edge}. By eliminating raw data transfer, FL overcomes satellite bandwidth limitations while preserving privacy. As 6G evolves, FL will underpin mission-critical applications from autonomous transportation to global environmental monitoring 

\subsection{Model Training and Deployment}

In machine learning, model compression is a strategy that minimizes performance loss while reducing the size and computing complexity of a learned model.

The process of merging separate models or model components into a bigger, more complete model is known as model aggregation. This can entail joining distinct modules with pre-established input and output ports to produce a more intricate model that is simpler to control and comprehend.

Model partitioning's primary objective is to enable parallel computing, which speeds up the simulation of big and complicated models by breaking them up into smaller segments that can fit within the set time-step constraints of real-time simulations.

\section{Conclusion}
TN-NTN 5G Advanced/6G IoT cloud- native and security fragmented security environment delivers AI native Federated DL/ML-powered visibility, prevention, and enforcement in a single platform.
Federated deep learning enables low-latency, secure intelligence at the edge, revolutionizing TN-NTN 5G Advanced/6G IoT. FL addresses satellite bandwidth restrictions while maintaining privacy by eliminating the transport of raw data.
key enablers like as network slicing, edge computing, AI-driven automation, and API-based capability exposing greatly increase the threat surface and raise the possibility of cyberattacks.
 5G-Advanced must have strong identity management, slice isolation, edge-level data protection, and secure orchestration of virtualized network operations in order to function in highly dynamic, programmable, multi-tenant environments.
Strict security guarantees are also necessary for enterprise and mission-critical use cases in order to comply with legal requirements and preserve cross-sector confidence.
FL will support mission-critical applications such as global environmental monitoring and autonomous mobility as 6G develops. 6G networks will use AI-native architecture that integrates intelligence across all tiers. The end-to-end (E2E) framework for AI-native Cloud networks in 5G advanced /6G will include AI on devices, RAN, core, IoT, and cybersecurity.

\bibliographystyle{ieeetr}

\end{document}